\documentclass[conference]{IEEEtran}
\IEEEoverridecommandlockouts

\makeatletter
\newcommand{\linebreakand}{%
 \end{@IEEEauthorhalign}
 \hfill\mbox{}\par
 \mbox{}\hfill\begin{@IEEEauthorhalign}
}
\makeatother

\usepackage{stfloats}
\usepackage{paralist}
\usepackage{graphicx}
\usepackage{verbatim}
\usepackage[bookmarks=false]{hyperref}
\hypersetup{
colorlinks = true, 
linkcolor=black, 
citecolor=black, 
urlcolor=black 
}

\usepackage[T1]{fontenc}
\usepackage[utf8]{inputenc}

\begin{document}
\title{Towards Quantum Software Requirements Engineering
}

\author{

\IEEEauthorblockN{Tao Yue}
\IEEEauthorblockA{
\textit{Simula Research Laboratory}\\
Oslo, Norway \\
tao@simula.no}
\and
\IEEEauthorblockN{Shaukat Ali}
\IEEEauthorblockA{
\textit{Simula Research Laboratory and}\\
\textit{Oslo Metropolitan University}\\
Oslo, Norway \\
shaukat@simula.no
}
\and
\IEEEauthorblockN{Paolo Arcaini}
\IEEEauthorblockA{
\textit{National Institute of Informatics}\\
Tokyo, Japan \\
arcaini@nii.ac.jp}

}

\maketitle
\begin{abstract}
Quantum software engineering (QSE) is receiving increasing attention, as evidenced by increasing publications on topics, e.g., quantum software modeling, testing, and debugging. However, in the literature, quantum software requirements engineering (QSRE) is still a software engineering area that is relatively less investigated. To this end, in this paper, we provide an initial set of thoughts about how requirements engineering for quantum software might differ from that for classical software after making an effort to map classical requirements classifications (e.g., functional and extra-functional requirements) into the context of quantum software. Moreover, we provide discussions on various aspects of QSRE that deserve attention from the quantum software engineering community.
\end{abstract}


\section{Introduction}
Quantum software engineering (QSE)~\cite{CACM2022,Piattini2022}, as classical software engineering, is expected to focus on various phases of quantum software development, including requirements engineering, design and modeling, testing, and debugging. Various studies have been conducted in the literature regarding most of these phases. However, as reported in~\cite{zhao2020quantum,CACM2022}, the requirements engineering phase remains relatively untouched. Only a few preliminary works exist on requirements engineering~\cite{QuantumUsecase,saraiva2021non}.

Requirements engineering, like in the classical context, if not conducted properly, will build incorrect quantum software and cause high costs in fixing it once problems are discovered in later phases of quantum software development. Thus, this paper focuses on quantum software requirements engineering (QSRE). In particular, we highlight the key aspects of QSRE that differentiate itself from the classical domain. To illustrate differences, we also present a motivating example of financial risk management. Moreover, we shed light on how typical requirements engineering will be impacted due to the quantum context and suggest key following activities.

\section{Motivating Example} \label{sec:motivatingexample} 
We will use the motivating example of credit risk analysis with quantum algorithms from Qiskit ~\cite{QiskitTutorialFinance}. Detailed information about the algorithm is published in~\cite{riskanayliswithqc,qriskanalysis}. The proposed quantum algorithm is more efficient than its equivalent classical implementations, such as using Monte Carlo simulations on classical computers.
We calculate two key risk measures, i.e., {\it Value at Risk} (VaR) and {\it Conditional Value at Risk} (CVaR). Key requirements of the risk estimation, including the calculation of these two risk measures (i.e., functional requirements), are shown in Figure~\ref{fig:requirementsDiagram}. In addition, we present extra-functional requirements specific to quantum computing, e.g., estimating the number of gates and (ancilla) qubits. Moreover, we show hardware constraints such as the limited number of qubits and limited depth of circuits.

Figure~\ref{fig:usecase}~(a) present a use case diagram including actor \textit{Credit Analyst} responsible for managing risk in finance, as illustrated with use case \textit{Manage risk in finance with quantum}. This use case includes use cases \textit{Determine Var} and \textit{Determine CVaR}. Also, for calculating VaR or CVaR, a credit analyst needs to define the confidence level, captured with use case \textit{Define the confidence level}.
In Figure~\ref{fig:usecase}~(b), we use the use case diagram notation to illustrate the main functionalities of a quantum expert applying the Amplitude Estimation~\cite{riskanayliswithqc,qriskanalysis} algorithm for calculating VaR and CVaR.

\section{Quantum Software Requirements Engineering}\label{sec:re4qss}

\subsection{Stakeholders}\label{subsec:stakeholder}
The ISO/IEC/IEEE 15288 standard defines stakeholders as: ``\textit{Individual or organization having a right, share, claim, or interest in a system or in its possession of characteristics that meet their needs and expectations}''~\cite{ISO15288}. Identifying stakeholders and their requirements is a crucial activity in requirements engineering. When building quantum software systems, stakeholders are the same as in the classical context. For example, in our example, stakeholders related to the development of the quantum risk management system include credit analysts (domain experts), borrowers (customers), banks, and software developers, all having different concerns on various aspects, including functionality, ease of use, price, and performance.

\subsection{Requirements classifications}\label{subsec:requirementstype}
Requirements are commonly classified into \textit{functional} and \textit{extra-functional}  (Section~\ref{subsubsec:FR4QSS}). A further classification specific to QSRE is whether requirements are related to the quantum or the classical part (Section~\ref{subsubsec:qReqandcReq}) of the system.

\subsubsection{Functional requirements and extra-functional requirements}\label{subsubsec:FR4QSS}
\begin{figure}[!t]
\centering
\includegraphics[width=1.0\linewidth]{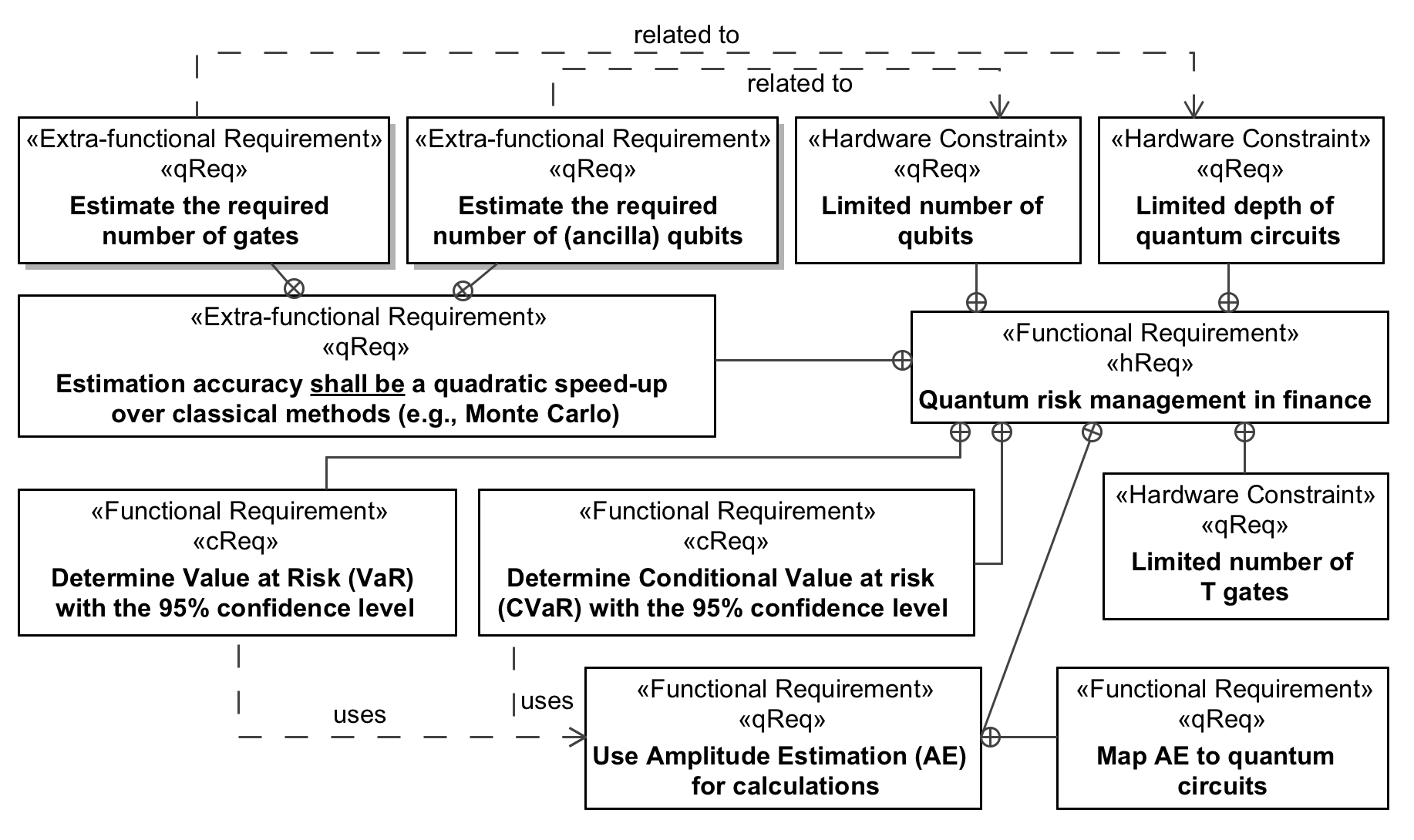}
\caption{Finance application for credit risk analysis -- key requirements, in the SysML requirements diagram notation.
Stereotypes \guillemotleft qReq\guillemotright, \guillemotleft cReq\guillemotright, and \guillemotleft hReq\guillemotright \space are applied to distinguish quantum requirements, classical requirements and the hybrid of both, respectively. Stereotypes \guillemotleft Functional Requirement\guillemotright \space and \guillemotleft Extra-functional Requirement\guillemotright \space distinguish functional and extra-functional requirements.
}
\label{fig:requirementsDiagram}
\end{figure}

\emph{Functional requirements} are related to the functionality that a quantum software system is expected to provide. 
For instance, the functional requirements of our example are indicated with \guillemotleft Functional Requirement\guillemotright, such as \textit{Determine Value at Risk (VaR) with the 95\% of confidence level} (see Figure~\ref{fig:requirementsDiagram}).
Identifying functional requirements for quantum software shall be the same as for the classical one.

SEBoK defines \emph{non-functional requirements} (also commonly named \emph{extra-functional}) as ``\textit{Quality attributes or characteristics that are desired in a system, that define how a system is supposed to be}''~\cite{QSLC}. These attributes vary from one system to another. For instance, safety requirements (i.e., one type of extra-functional requirements) will only apply to a safety-critical system. All the relevant extra-functional requirements from classical software systems generally apply to quantum software systems. However, there are additional requirements. For instance, Figure~\ref{fig:requirementsDiagram} shows three extra-functional requirements: \textit{Estimation accuracy shall be a quadratic speed-up over classical methods (e.g., Monte Carlo)}, which is further decomposed into another two extra-functional requirements on estimating the required numbers of gates and (ancilla) qubits. These two requirements relate to the hardware constraints: \textit{Limited number of qubits} and \textit{Limited depth of quantum circuits}. Identifying and realizing these extra-functional requirements require knowledge of quantum computing. The good news is that such requirements are common across various quantum software applications, implying that they can be reused and that common solutions can be proposed to address them. We would also like to point it out that Saraiva et al.~\cite{saraiva2021non} have already identified such five common extra-functional requirements. Moreover, such extra-functional requirements might need to be elicited step-wise, as their elicitation depends on identifying other requirements. Ideally, when available in the future, an actionable requirements elicitation process could clearly guide users through all required activities.

\subsubsection{Quantum requirements vs. classical requirements}\label{subsubsec:qReqandcReq}

\begin{figure}[!tb]
\centering
\includegraphics[width=0.9\linewidth]{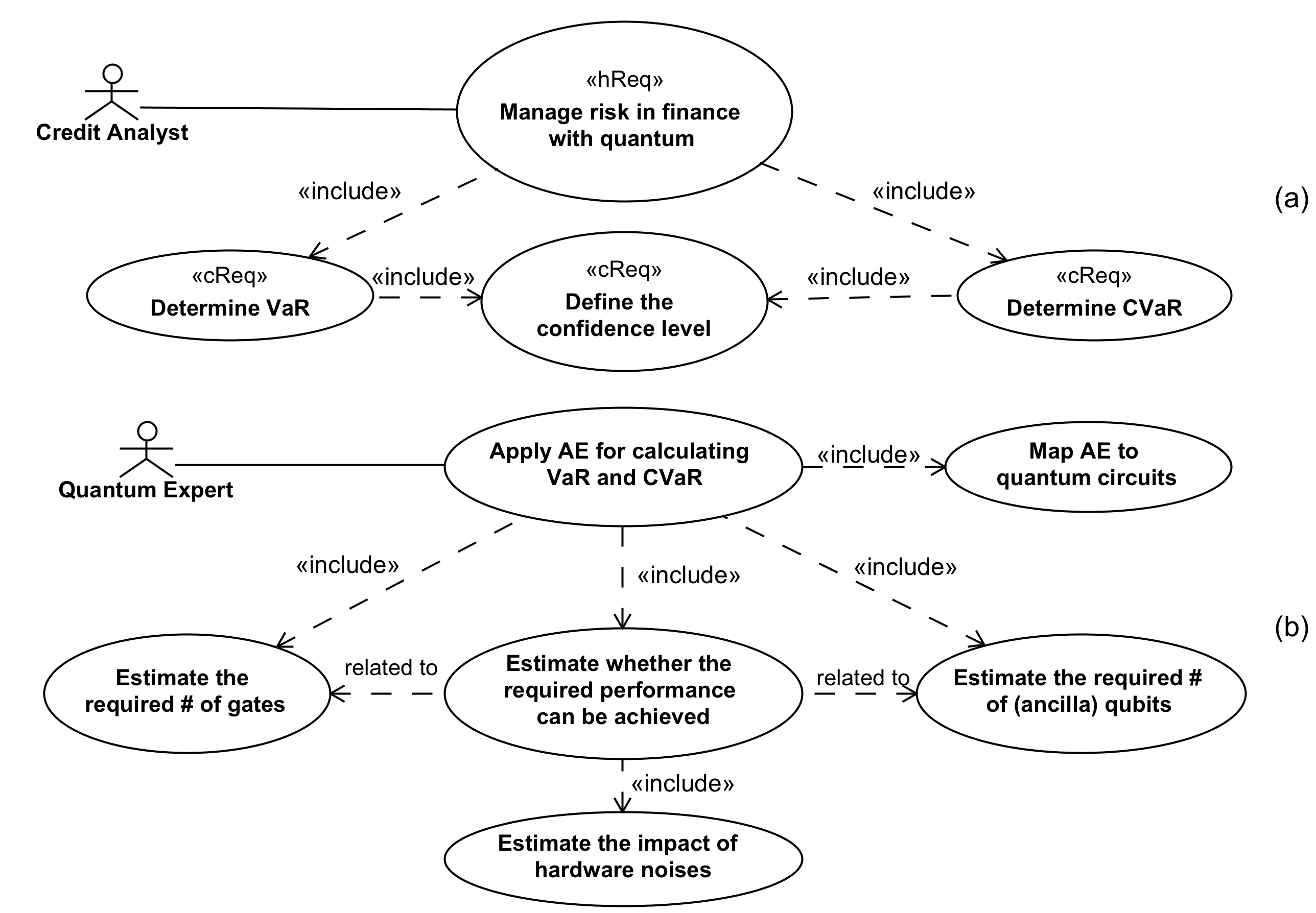}
\caption{(a) Application for credit risk analysis -- key use cases. 
(b) Key functionalities of realizing \textit{Determine VaR} and \textit{Determine CVaR} in (a).}
\label{fig:usecase}
\end{figure}

It is crucial to distinguish requirements that should be addressed with classical computers and those to be addressed with quantum computers. Moreover, there should be high-level requirements that are hybrid. For instance, Figure~\ref{fig:requirementsDiagram} defines three stereotypes \guillemotleft qReq\guillemotright, \guillemotleft cReq\guillemotright, \space and \guillemotleft hReq\guillemotright \space to distinguish requirements that need to be addressed in the classical, quantum, or hybrid manner, respectively. Doing so is essential, as mentioned by Weder et al.~\cite{QSLC,Weder2022}; indeed, the first phase of their proposed quantum software lifecycle is about performing quantum-classical splitting. Requirements engineering, especially requirements analysis, and optimization, is typically performed in this phase to decide which parts of a targeted problem need to be solved on a quantum computer and which parts go to classical hardware. Consequently, requirements specification and modeling solutions should provide mechanisms to support the problem separation into classical and quantum parts. We explain this idea by applying three stereotypes to use case modeling (see Figure~\ref{fig:usecase}).

\subsection{Specific extra-functional concerns} \label{subsubsec:bilities}

\subsubsection{Portability}
Near future quantum computers will be built with different hardware technologies; thus, portability will remain a key requirement to be captured. For example, a quantum software system for our example (i.e., credit risk analysis) may need to be deployed to different quantum computers. Moreover, in the near future, various quantum computing resources and classical computing will be pooled so that more computations can be performed jointly. Thus, converting a problem into a set of requirements, where requirements shall be addressed with different types of quantum computers and classical computers, is needed.

\subsubsection{Performance}
Performance requirements over classical implementation are essential to justify the need for quantum computing. For example, our example requires that the estimation accuracy be a quadratic speed-up over classical methods (see Figure~\ref{fig:requirementsDiagram}). Such requirements may consider other requirements, e.g., the estimation accuracy depends on the number of gates and the number of (ancilla) qubits that need to be estimated at the requirements level to check whether or not the expected quadratic speed-up on the estimation accuracy can be achieved. These are two additional requirements. Such requirements are common across quantum software, as, currently and in the near future, the capabilities of quantum computers are limited. Thus, deciding early on whether available resources can achieve the expected performance requirements and, if yes, with which margin is important.

\subsubsection{Reliability}
Currently, hardware errors affect the reliability of their computations and consequently constrain how quantum computers should be used and how quantum software should be designed. For instance, performing a reset after running several calculations for a period of time might be needed; this means that a quantum algorithm might not be run for a long time~\cite{BoostingTensorflow}. Thus, when identifying requirements of quantum software, it is essential to identify reliability requirements and associated constraints, especially considering the impact of hardware errors on the reliability of quantum software systems. Decisions such as introducing Quantum Error Correction~\cite{QEC} (which requires additional quantum resources) or other fault tolerance mechanisms might be needed early in the quantum software development lifecycle.

\subsubsection{Scalability}
Current quantum computers support a limited number of qubits, i.e., resources are scarce and expensive. Therefore, scalability requirements are carefully considered while designing quantum software. For instance, as discussed in~\cite{qriskanalysis}, in the context of quantum risk analysis (our motivating example), based on the results of the authors' investigation, more qubits are needed to model more realistic scenarios, thereby achieving practically meaningful advantages over Monte Carlo simulations, which represent state of the art in risk management. Moreover, scalability requirements (e.g., on the number of parameters and constraints expected to be handled in the risk analysis) should be carefully defined such that they can be satisfied with a limited depth of the quantum circuit to mitigate the impact of decoherence, with limited use of two-qubit gates (e.g., CNOT gates) to reduce the effect of crosstalk, and so on, which can be ensured with more powerful quantum computers, dedicated error mitigation mechanisms, and even carefully-designed quantum algorithms.

\subsubsection{Maintainability}
Like classical software, quantum software will require maintainability. Given that, as expected, quantum hardware will continue to evolve, existing quantum software needs to be updated (in some cases) to deal with the hardware changes. For example, with the decreased hardware error rates provided by the latest technological advancements, error handing mechanisms in quantum software systems must be updated to improve performance and reduce the cost of additional error correction. Thus, quantum software systems shall identify and capture such maintainability requirements.

\subsubsection{Reusability}
Like classical software, the reusability of quantum software is essential to be easily reused across different systems. Thus, such requirements shall be captured during requirements engineering. However, some specific requirements related to quantum software shall be explicitly captured. For instance, quantum software is often built as hybrid software. Therefore, having tight coupling between the two parts would reduce the reusability of quantum software. Instead, the high cohesion of the quantum software part is expected to enable more reusability.

\section{Discussions and Suggestions}\label{sec:discussion}

Requirements elicitation elicits software requirements, i.e., quantum software in our context. Given that, in this phase, we investigate \textit{what} problem a quantum software should solve rather than \textit{how} the software should be implemented to address this problem, the requirements engineering for quantum software shall remain similar to the classical one. For instance, identifying stakeholders and defining system boundaries remain the same. However, one difference might be in checking whether it is needed to solve a problem that has been solved in the classical world with quantum, especially considering the known limitations of quantum computing. For example, in our running example (see Figure~\ref{fig:requirementsDiagram}), we need to consider requirements specific to the quantum domain, such as the required number of qubits (i.e., a hardware constraint). Regarding stakeholders, there remain similarities between the classical and quantum requirements elicitation. For instance, a possible stakeholder in our example is the credit analyst, which would remain the same as in the classical domain. Existing methods, such as interviews and prototyping for requirements elicitation, are also expected to be largely similar.

Functional and non-functional requirements are typically specified during requirements specification at various formalization degrees, ranging from informal natural language specifications to fully formal specifications. Examples include semi-formal notations such as use cases and entity-relations diagrams or formal notations such as Hoare logic~\cite{zhou2019applied}. Requirements specifications for quantum software will be changed to accommodate concepts related to quantum software. For example, when using use case diagrams, as shown in our example, it is helpful to distinguish use cases from the classical world, the quantum world, and the mix of the two. Moreover, when specifying requirements with modeling notations (e.g., SysML requirements diagram), they need to be extended to capture novel concepts from quantum software. Finally, formal methods are also relevant to investigate for specifying quantum software requirements as surveyed in~\cite{formalmethodsqp}. Nonetheless, such methods are also quite early in their stage of development~\cite{formalmethodsqp}.

Requirements \textit{verification} of quantum software has received less attention; when considering formal methods, only preliminary tools and methods are available as discussed in~\cite{formalmethodsqp}. Moreover, the survey discusses the need for new methods for formal verification for complex quantum software. Requirements validation via automated testing is getting popular in the software engineering community, with several new works being published (e.g., \cite{genTestsQPSSBSE2021,honarvar2020property,ourICST2021,quantumCombTestQRS21,Muskit,10.1145/3528230.3529186}).
Nonetheless, as discussed in~\cite{MiranskyyICSE19}, many testing challenges remain unaddressed. Finally, the classical verification and validation methods, e.g., inspection and walk-through, apply to some extent to quantum software requirements.

Based on our investigation, we recommend the following:

(1) Carefully consider separating parts of the problem that should be addressed in the classical world and those on quantum computers; (2) Identify and specify requirements related to various constraints, especially those about quantum hardware. Realizing these requirements depends on available and realistic quantum computing resources and explicitly specifying such requirements support requirements analysis on the feasibility of the realization; (3) Identify existing quantum algorithms that could be incorporated. Selecting which quantum algorithms to use is a decision that might need to be made at the early stage, as the availability and capability of such quantum algorithms have an impact on the quantum part of the realization of certain extra-functional requirements; (4) Based on the identified and specified requirements, requirements analysis might be needed to identify key factors (e.g., selection of quantum algorithms, determining quantum hardware resources, assessing the feasibility of satisfying extra-functional requirements) that have a significant impact on the development of quantum software, and potential trade-offs among these factors. Doing so is expected to effectively support decision-making on selecting quantum hardware resources; (5) Identify requirements whose realization strongly depends on constantly emerging quantum algorithms and advanced quantum computers. Doing so is necessary because as soon as more advanced quantum algorithms or quantum computers are available, such requirements could be realized (if not possible before) or realized better. Also, decisions made regarding the satisfaction of certain requirements (e.g., the required number of gates) and rationales behind these decisions are highly recommended to be recorded.

\section{Conclusions and Future Work}\label{sec:conclusion}
Requirements engineering (RE) for quantum software has gotten less attention than other phases, such as quantum software testing. Thus, we present some ideas on how RE for quantum software will differ from the classical counterpart. For instance, what will be the key differences for extra-functional requirements? Finally, we discussed how various steps in RE, such as requirements elicitation, specification, verification, and validation, will be impacted, including developing requirements specification/modeling, analyses, verification, and validation methods, with tool support, for supporting quantum software development at the RE phase.

\bibliographystyle{IEEEtran}
\bibliography{IEEEabrv,References.bib}

\end{document}